\documentclass[twocolumn]{revtex4-2}
\usepackage[english]{babel}
\usepackage[T1]{fontenc}
\usepackage[colorlinks=true,urlcolor=blue]{hyperref}
\usepackage{mathtools}
\usepackage{xcolor}
\usepackage{bm}

\definecolor{red}{rgb}{0.75,0,0}
\definecolor{blue}{rgb}{0,0,0.75}
\definecolor{green}{rgb}{0,0.5,0}

\begin{document}

\title{Theory of cellular homochirality and trait evolution in flocking systems}
\author{Ludwig A. Hoffmann}
\author{Luca Giomi}
\email{giomi@lorentz.leidenuniv.nl}
\affiliation{Instituut-Lorentz, Universiteit Leiden, P.O. Box 9506, 2300 RA Leiden, The Netherlands}
\date{\today}

\begin{abstract}
Chirality is a feature of many biological systems and much research has been focused on understanding the origin and implications of this property. Famously, sugars and amino acids found in nature are homochiral, i.e., chiral symmetry is broken and only one of the two possible chiral states is ever observed. Certain types of cells show chiral behavior, too.
Understanding the origin of cellular chirality and its effect on tissues and cellular dynamics is still an open problem and subject to much (recent) research, e.g., in the context of drosophila morphogenesis.
Here, we develop a simple model to describe the possible origin of homochirality in cells.
Combining the Vicsek model for collective behavior with the model of Jafarpour {\em et al.}~\cite{jafarpour2015}, developed to describe the emergence of molecular homochirality, we investigate how a homochiral state might have evolved in cells from an initially symmetric state without any mechanisms that explicitly break chiral symmetry.
We investigate the transition to homochirality and show how the ``openness'' of the system as well as noise determine if and when a globally homochiral state is reached. We discuss how our model can be applied to the evolution of traits in flocking systems in general, or to study systems consisting of multiple interacting species.
\end{abstract}

\maketitle

\section{\label{sec:introduction}Introduction}

Chirality -- i.e. the property of an object to differ from its mirror image -- is a common feature of many biological systems: from amino acids up to biopolymers, cells and fully developed vertebrate and invertebrate organisms~\cite{lenz2008,alberts2015,nelson2020}. Most biomolecules, for instance, are either left- or right-handed (L or R for brevity), despite the reactions originating them not favoring either one of the two handedness. Thus most amino acids are left-handed whereas R-molecules are predominant among sugars. These small chiral molecules, in turn, serve as building blocks of larger biopolymers, thereby providing the molecular basis for a hierarchical inheritance of chirality by larger structures. 

Because of its ubiquitous in biology, the origin and the relevance of chirality has been the subject of insightful research for decades; see, e.g., Refs.~\cite{frank1953,mason1988,blackmond2010,inaki2016,jafarpour2017} and references therein. The function of a specific handedness, if any, and the mechanism leading to chiral symmetry breaking, in particular, are still hotly debated. While chirality appears crucial in small biomolecules, where it was found to be linked to the function of e.g. proteins~\cite{inaki2016}, it is not always clear if and how chirality plays a role in larger structures. Thus, while certainly LR asymmetry is instrumental to the mechanics of flagella in sperm cells~\cite{gray1955,gaffney2011} and bacteria~\cite{lauga2016}, its occurrence and role is not equally obvious in eukaryotic cells. Yet, an increasingly large body of experimental evidence has recently started to indicate that, even in this case, chirality could serve specific biophysical functions in both  unicellular \cite{inaki2016} and multicellular systems. For example, cell chirality has been shown to influence the morphogenesis of Drosophila \cite{taniguchi2011}, snails \cite{davison2016}, \textit{C. elegans} \cite{pohl2010}, and mammalian cells \cite{xu2007,wan2011,worley2015,chin2018}. {\em In vitro}, a particularly compelling example chirality and its effect in multicellular eukaryotes was reported by Duclos {\em et al}. in confined layers of spindle-like RPE1 and C2C12 cells~\cite{duclos2018}. Because of the elongated shape of their constituents, these cellular fluids are unstable to spontaneous bending and flow, but, unlike in achiral active nematics, where the loss of hydrodynamic stability takes place via a standard pitchfork bifurcation~\cite{voituriez2005}, the active stresses fueling the instability here feature an additional contribution that explicitly breaks LR symmetry~\cite{hoffmann2020}, thus providing the cellular fluid with a specific and reproducible chirality. 

Now, while able to account for the post-transitional scenario in confined nematic cell monolayers -- from the basic phenomenology~\cite{duclos2018,hoffmann2020}, to the more subtle aspects, such as the existence of topological edge currents observed in the chaotic regime~\cite{yashunsky2022} -- these chiral active stresses must originates, at the microscopic scale, from a spontaneous break down of LR symmetry, which is not captured by any of the current hydrodynamic theories of nematic cell monolayers. In this article we bridge this gap by investigating the microscopic origin of chirality in cell monolayers, and how collective motion influences the occurrence of a spatially homogeneous chirality -- also referred to as {\em homochirality} -- at the macroscopic scale. Our model follows a classic approach to chiral symmetry breaking, pioneered by Frank~\cite{frank1953} and recently extended by Jafarpour {\em et al.}, to explain the emergence of homochiral states in racemic mixtures of L and R molecules~\cite{jafarpour2015, jafarpour2017}. Starting from a minimal model based on two reaction equations, we show that noise breaks LR symmetry already at the cellular scale and the system can reach a homochiral state from an initially racemic one. Compared to the analysis by Jafarpour {\em et al.}, where the molecules are assumed to undergo passive diffusion, our model mesenchymal-like cells consist of self-propelled particles with aligning interactions, and whose dynamics is governed by the classic Vicsek model~\cite{vicsek1995}. This results in markedly different spatiotemporal dynamics. We numerically investigate if and how the presence of activity and alignment interactions influences the transition to homochirality from a racemic state. We find that, for specific choices of parameters, the system is guaranteed to reach a homochiral state in a finite time. While the system is in a mixed state, hence away from homochirality, we find large fluctuations of the number density and the local chirality. Furthermore, we observe that like-chiral cells are more strongly correlated than cells of opposite chirality, even though there is no explicit interaction term favoring one over the other. Finally, we find that the time a given system takes to transition to homochirality follows a long-tail distribution, with mean and standard deviation being of the same order of magnitude.

In the following section we first describe our model in more detail. In Sec.~\ref{sec:results}, on the other hand, we identify the conditions for which the system transitions from a racemic to an homochiral state and further investigate the statistics of spatial fluctuations and heterogeneities. Afterwards, we consider in more detail the transition to homochirality, and how the transition time depends on several of the parameters of the model.

\section{\label{sec:model}Model}

\begin{figure*}
\centering
\includegraphics[width=0.75\textwidth]{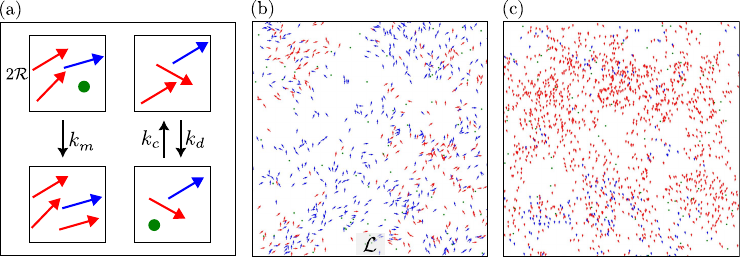}
\caption{{\bfseries Model.} (a) A schematic of the reactions defined in Eqs.~\eqref{eq:Reactions}. To perform the reactions we subdivide the total system into square boxes of  height $2\mathcal{R}$. Left-chiral particles (blue arrow) and right-chiral particles (red arrow) can divide with the rate $k_m$ through the consumption of nutrients (green dot). Here a left-chiral particle divides. Furthermore, they die with the rate $k_d$ and are replaced by a nutrient, or the nutrient ``spontaneously transforms'' into a particle with rate $k_c$. Here a left-chiral particle undergoes these reactions. See the main text for a proper interpretation of the rates. (b) Snapshot of the simulations carried out in a square system of height $\mathcal{L}$ at density $\rho = 2$ and noise $\eta = 0.3$. Again, right-chiral particles are represented by blue arrows while left-chiral particles are represented by red arrows. (c) A snapshot taken for higher densities ($\rho = 4$) and lower noise ($\eta = 0.1$).}
\label{fig:model}
\end{figure*}

Our model consists of two distinct processes that are minimally coupled: one describing cell division and death, the other the spatial dynamics, i.e., the collective motion of the cellular flock. To account for cell division and death, we adopt a version of a stochastic model originally introduced by Frank~\cite{frank1953} and recently expanded by Jafarpour {\em et al.}~\cite{jafarpour2015, jafarpour2017}. This model assumes the existence of two possible chiralities -- i.e. L and R -- and a solvent -- denoted with S -- which can fill the space left by a cell after its death and replenish the cell layer with nutrients. These processes occurs by means of the following reactions: 
\begin{subequations}\label{eq:Reactions}
\begin{align}
{\rm R} + {\rm S} \stackrel{k_m}{\longrightarrow} 2 {\rm R}\;, \qquad &{\rm L} + {\rm S} \stackrel{k_m}{\longrightarrow} 2 {\rm L}\;, \label{eq:Reactions_Division} \\
{\rm R} \xrightleftharpoons[k_c]{k_d} {\rm S}\;, \qquad &{\rm L} \xrightleftharpoons[k_c]{k_d} {\rm S} \label{eq:Reactions_Death} \;.
\end{align}
\end{subequations}
The first two equations, Eqs.~\eqref{eq:Reactions_Division}, describe cell division: a cell with given chirality uptakes nutrients from the solvent and divides into two cells having the same chirality with a rate $k_m$. The second set of equations, Eqs.~\eqref{eq:Reactions_Death}, contains a forward and a backwards reaction. The former, occurring with rate $k_d$, accounts for the death of a cell and its extrusion from the monolayer (apoptosis), after which the void left by the cell is replenished with solvent. The backwards reaction, occurring with rate $k_c$, describes instead cells entering the monolayer from an external reservoir and replacing part of the solvent in the process. Thus, the reaction rate $k_c$ encodes the ``openness'' of the cell layer, that is its propensity to recruit cells form outside a specific region of interest or from another layer situated above or below. If the rate vanishes the monolayer is closed and no cells can enter, while if the rate is positive there is a non-vanishing flux of cells into the monolayer. Our model relies then on three non-trivial reaction rates: $k_m$ (cell division), $k_d$ (cell death), and $k_c$ (cell influx). Note also that all reactions are symmetric for left and right chirality, i.e., there is no explicit symmetry breaking. While undoubtedly simplistic when compared to the actual life cycle of a cell embedded in a monolayer, these reactions allows us to account for two fundamental cellular processes, such as division and apoptosis, while rendering the problem tractable, by guaranteeing that the total number of cells and solvent particles, i.e. $N=N_R+N_L+N_S$, is conserved. Moreover, while originating in the realm of chiral systems, Eqs.~\eqref{eq:Reactions} could potentially describe the inheritance of {\em any} trait in a community of cells (or other active particles) that divide, die, and enter and exit an open environment. Some of our results can be thought as generic of cellular flocks, whether chiral or achiral. 

Cellular motion is described in terms of the classic Vicsek model~\cite{vicsek1995} (see also Refs.~\cite{vicsek2012,ginelli2016,chate2020} for reviews). Each cell is characterized by a position $\bm{r}_{i}$ and a velocity $\bm{v}_{i}=v_{0}(\cos\theta_{i},\sin\theta_{i})$, with $v_{0}$ a constant, whose evolution in time is governed by the following set of recursion relations
\begin{subequations}\label{eq:vicsek}
\begin{align}
\bm{r}_i(t + \Delta t) &= \bm{r}_i(t) +  \bm{v}(t)\Delta t\;,\\
\theta_i(t + \Delta t) &= {\rm Arg} \left[\sum_j C_{ij}(t) \bm{v}_j(t) \right] + \lambda\Omega+\xi_i(t)\;.
\end{align}
\end{subequations}
Here, $C_{ij}(t)$ is the connectivity matrix whose entries are $C_{ij}(t) = 1$ if $|\bm{r}_i(t) - \bm{r}_j(t)| < \mathcal{R}$, with $\mathcal{R}$ a constant interaction radius, and $C_{ij}(t) = 0$ if $|\bm{r}_i(t) - \bm{r}_j(t)| > \mathcal{R}$. The second term in Eq.~(\ref{eq:vicsek}b) describes a deterministic rotation of a given particle, with $\Omega$ the magnitude, while $\lambda$ sets the orientation of the rotation (thus, $\lambda = 1$ for left-chiral cells, and $\lambda = -1$ for right-chiral cells, respectively). This couples the chirality of a given cell to its motion in space. To keep the model as simple as possible we will in the following set $\Omega = 0$, thus not include any deterministic rotation. The equations of motion for L and R cells are then identical and chirality does not affect the spatial dynamics explicitly. We will briefly discuss the effect of a non-vanishing rotation rate in the Discussion section at the end of this article. Finally, $\xi_i(t)$ is a Gaussian-distributed random number with zero mean, $\langle \xi_{i} \rangle = 0$, and finite standard deviation, $\sqrt{\langle \xi_{i}^{2}\rangle}=\eta$. Thus, the first two terms on the right-hand side of Eq.~(\ref{eq:vicsek}b) aligns the direction of motion of the $i-$th cell with those of its neighbors, which in turn can process at the rate $\pm \Omega/\Delta t$ depending on the chirality of the cells. The third term, on the other hand, introduces a random rotation whose effect is to disturb such an alignment mechanism, thereby favoring isotropy across the flock. The relative importance of these two effects is determined by the constant $\eta$, which, in our construction, varies in the range $0 < \eta < 1$. If $\eta$ is sufficiently small and the density of particles is sufficiently large, the system described by Eqs.~\eqref{eq:vicsek} undergoes a discontinuous phase transition from a disordered to a flocking state, where all the agents persistently move in the same direction.

The reactions in Eqs.~\eqref{eq:Reactions} are implemented via the Gillespie algorithm \cite{gillespie1976,gillespie2007} and coupled with the dynamics described by Eqs.~\eqref{eq:vicsek} using the following strategy. After each time step of the Vicsek model we divide the total system into boxes of area $(2\mathcal{R})^2$ and in each of these boxes we run $m$ steps of the Gillespie algorithm. After updating the population in each box, we perform another time step of the Vicsek model. Our {\em in silico} cell monolayer inhabits a square box of size $\mathcal{L}$ with periodic boundary and, at $t=0$, consists solely of one L and one R cell, with random positions and orientations. At a given density $\rho$ there are then $N_S (t = 0) = \rho \mathcal{L}^2$ number of solvent particles. Thus, the total number of agents, including both cells and solvent particles, for a given density $\rho$ is given by $N = \rho \mathcal{L}^2 + 2$ which, as explained above, is conserved and constant in time by construction. We fix length scales by setting the interaction radius to unity, $\mathcal{R} = 1$, and time scales by setting the time step in the Vicsek model to unity, $\Delta t = 1$. In these units, we set $k_d = 10$ and define $\tilde{k}_m = k_m/k_d$ and $\tilde{k}_c = k_c/k_d$. In the following we will always work with the rescaled rates, but drop the tilde. The effect of varying the other model parameters will be investigated below. In Figs.~\ref{fig:model}b,c we show a snapshot of the simulations at different densities and values of $N_L/N_R$. While the global alignment in Fig.~\ref{fig:model}b is low, for higher values of density and lower values of noise almost all cells have the same orientation in Fig.~\ref{fig:model}c. We color L cells in red and R cells in blue. Note that the configuration shown in Fig.~\ref{fig:model}c is considerably closer to homochirality than that shown in Fig.~\ref{fig:model}b, with $N_L \gg N_R$. The question whether the density and flocking has an effect on the appearance of homochirality, or the mean time until this state is reached, will be discussed below.

\section{\label{sec:results}Results}

To explore how homochirality is progressively established across the cellular flocks described by Eqs.~\eqref{eq:Reactions} and \eqref{eq:vicsek}, we first investigate the effect of the rate $k_c$. As we will see, the system is guaranteed to reach a homochiral state only if this rate vanishes. Therefore, as we are interested in the transition to homochirality, we afterwards set $k_c = 0$ and instead investigate the time it takes a given system on average to reach the homochiral state. We investigate how varying different model parameters speeds up or slows down the transition time.
\begin{figure*}
\centering
\includegraphics[width=\textwidth]{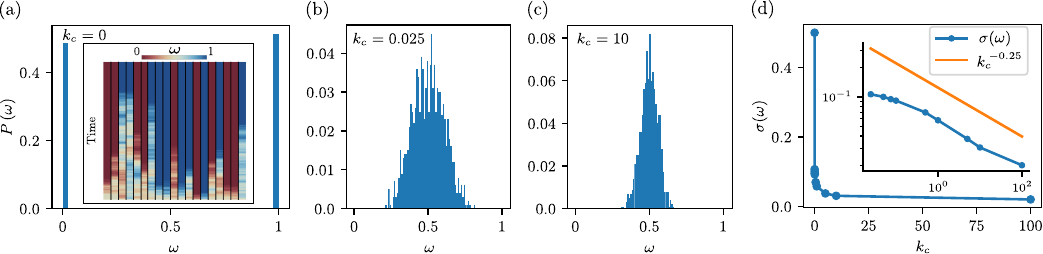}
\caption{{\bfseries Distributions of global order parameter.} (a) Probability distribution $P(\omega)$ of the order parameter $\omega$ for vanishing rate $k_c$. The system is guaranteed to evolve to a homochiral state, thus $\omega = 0$ and $\omega = 1$ both occur with a probability of about $50\%$. The inset represents the time evolution of the order parameter for twenty independent runs. Each of the columns is an independent run and time increases in positive $y$-direction. The color code is according to the legend at the top, i.e., $\omega = 0$ (homochirality of right particles) is red, $\omega = 1$ (homochirality of left particles) is blue, the racemic state $\omega = 0.5$ is white. Every systems is initialized in a state with $\omega = 0.5$. The order parameter can be seen to fluctuate in time but eventually all systems evolve to one of the two homochiral states. (b) Probability distribution of the order parameter for $k_c = 0.025$. (c) Distribution for $k_c = 1$. (d) Distribution for $k_c = 10$. To obtain each of the histograms we measure the order parameter of the system after a certain number of time steps $\Delta t$, where the system has on average reached a stationary state, and average over 1000 independent runs. (d) We plot the standard deviation $\sigma(\omega)$ of the distribution $P(\omega)$ as a function of the reaction rate $k_c$. We find that the curve is well approximated by a power law decay with exponent $-0.25$ (see inset for log-log plot of standard deviation $\sigma(\omega)$ over $k_c$). Simulation parameters: $k_m = 5$, $\rho = 2$, $\eta = 0.3$, $v_0 = 0.4$, $m = 2$.} 
\label{fig:distributions}
\end{figure*}

\subsection{Global properties of open cell layers}

As mentioned in Sec.~\ref{sec:model}, we interpret $k_c$ as the rate at which new cells of either chirality are introduced into the system, but not as a result of cell division. This can occur, for instance, in open cell monolayers, when a cell enters a specific region of interest, thereby replacing (consuming) previously present nutrients, or in multilayered structures, when a cell move from a layer to another: i.e., both processes cause a non-vanishing flux of cells into the system. This is assumed to happen equally likely for cells of either chirality, therefore guaranteeing that LR symmetry is not explicitly broken. Global chirality can be identified starting from the order parameter
\begin{equation}
\omega = \frac{N_L}{N_L + N_R} \in [0,1]\;,
\end{equation}
such that $\omega = 1$ if all particles are L, $\omega=0$ if all particles are R and $\omega=0.5$ in case of a racemic mixture of L and R. We now consider the probability distribution of $\omega$ after a given number of time steps of Eqs.~\eqref{eq:vicsek}, when the average of many independent runs has approximately reached a steady state. To obtain the probability distribution we record the order parameter at this time for $10^{3}$ independent simulations. The resulting distributions are shown in Fig.~\ref{fig:distributions} for some values of $k_c$. If $k_c$ vanishes we find a bimodal probability distribution which takes non-vanishing values only at the homochiral states $\omega = 0$ and $\omega = 1$, see Fig. \ref{fig:distributions}a. That is, regardless of the specific rates of division and apoptosis, the monolayer always converges within a {\em finite time} to a homochiral state, which is equally likely to be L or R. The time evolution of the order parameter for some of the runs is presented in the inset. As can be seen, the order parameter heavily fluctuates initially, but once a system has evolved into an homochiral state it remains in this state. This reflects the fact that, for $k_c = 0$, once $N_L = 0$, new L cells cannot be created from the reactions Eqs. \eqref{eq:Reactions}. The only reactions occurring in the case that $N_L = 0$ are cell division and death of R cells. Similarly for $N_R = 0$. Thus, the homochiral states are a fixed point of the reactions if $k_c = 0$.

If the creation rate is finite, however, the system is not guaranteed to reach a homochiral state. Indeed, we find that already for small $k_c$  values the probability distribution changes dramatically, with the distribution being peaked at the racemic state $\omega = 0.5$ (see Fig. \ref{fig:distributions}b for $k_c = 0.025$) and the monolayer never converging to a homochiral state. As $k_c$ increases, the width of the probability distribution decreases rapidly and for large $k_c$ values the distribution is sharply peaked around $\omega = 0.5$ (Fig. \ref{fig:distributions}c for $k_c = 10$). To quantify this behavior we computed the standard deviation $\sigma(\omega)$ of the distributions as a function of the rate $k_c$ over four orders of magnitude. We find that approximately $\sigma(\omega) \sim k_c^{-1/4}$, Fig. \ref{fig:distributions}d. Lastly, note that the distribution is symmetric for all $k_c$ values, reflecting that none of the mechanisms entailed by Eqs.~\eqref{eq:Reactions} and \eqref{eq:vicsek} explicitly breaks LR symmetry. The average order parameter is, therefore, always $\langle \omega \rangle = 0.5$.

\subsection{Spatial fluctuations}

In the previous section we discussed the probability distribution of the chiral order parameter $\omega$ across the entire system. As already evident from the simulations of the snapshots in Figs.~\ref{fig:model}b,c, neither $\omega$ nor the cell number density are uniform across the monolayer, but vary greatly in space. Such an inhomogeneity originates from the anisotropy introduced through the alignment interaction and is enhanced by the large density fluctuations that characterize the Vicsek model.

\begin{figure}[t]
\centering
\includegraphics[width=\columnwidth]{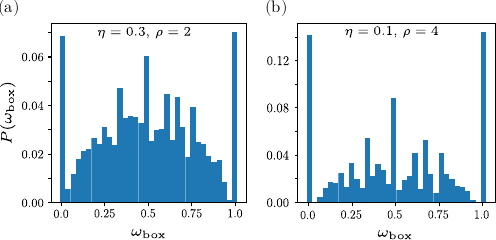}
\caption{{\bfseries Distributions of local order parameter.} We measure the local order parameter $\omega_\text{box}$, that is the order parameter in each of the square boxes of height $2\mathcal{R}$. Per system there are $(\mathcal{L}/2\mathcal{R})^2 = 100$ boxes. To obtain the histograms we show here we average over 1000 independent runs. If a box is empty the order parameter is not well defined and we do not include it in the histogram. (a) The probability distribution for noise $\eta = 0.3$ and density $\rho = 2$. (b) The probability distribution for noise $\eta = 0.1$ and density $\rho = 4$. Simulation parameters: $k_c = 0.025$, $k_m = 5$, $v_0 = 0.4$, $m = 2$.}
\label{fig:dist_local_op}
\end{figure}

\begin{figure*}[t]
\centering
\includegraphics[width=\textwidth]{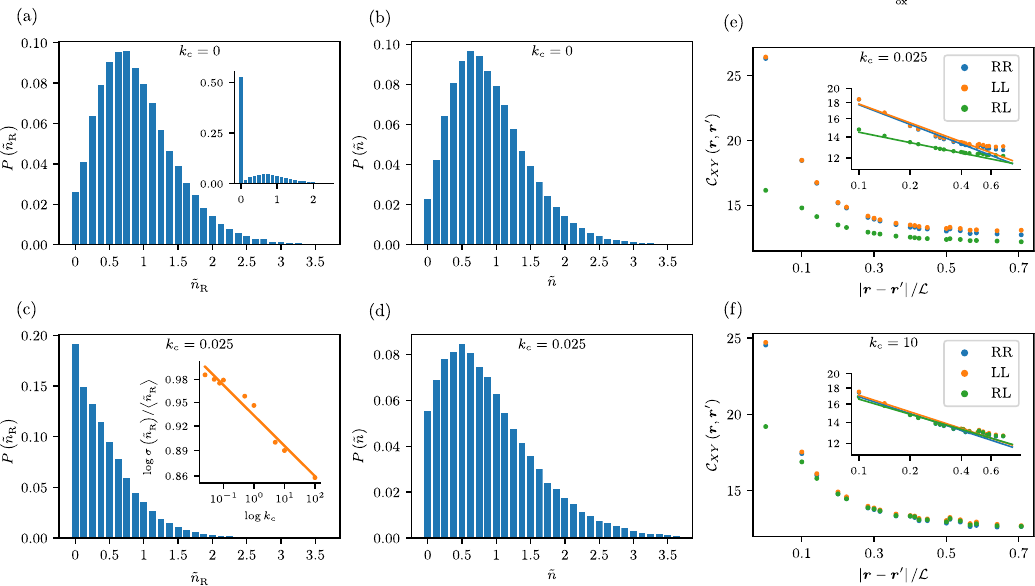}
\caption{{\bfseries Fluctuations and correlations of particles density.} (a) We show the probability distribution to find a number of $\tilde{n}_R$ right-chiral particles in a box for $k_c = 0$. Here, $\tilde{n}_R$ is not the total number of particles but the ratio of number of particles over the average number of particles one would expect in a box in a homogeneous system (see main text). Thus for $\tilde{n}_R < 1$ the number of particles in a box is smaller than expected in a homogeneous system while for  $\tilde{n}_R > 1$ it is greater. The probability distribution shown here is obtained by ignoring/subtracting the cases where the system has evolved to a left-homochiral state. The inset shows the distribution if these cases are not subtracted. The large probability of finding a box without right-chiral particle reflects the fact that the system evolves in half of the runs to a left-homochiral where no right-chiral particles are present. (b) The probability to find a given relative total number of particles in a box. (c) The same probability distribution as in (a) but for $k_c = 0.025$ and without subtracting the cases of left homochirality. This is because, as was shown in Fig.~\ref{fig:distributions}, the system virtually never evolves to a homochiral state and the systems are always mixed. The inset shows the depends of the ratio of standard deviation over mean for the probability distribution $P(\tilde{n}_R)$ as a function of the rate $k_c$. The best fit is found to be $\sim k_c^{0.017}$. (d) The same distribution as in (b) but for $k_c = 0.025$. (e) The correlation function between right-chiral particles (RR, blue), left-chiral particles (LL, orange), and between left- and right-chiral particles (RL, green). The $x$-axis is distances measured in terms of the system size $\mathcal{L}$. The inset shows the same data in a log-log plot with the best fit of the linear region (in the log-log plot), namely $0.1 < |\bm{r} - \bm{r}'|/\mathcal{L} < 0.5.$ (f) The same data but for $k_c = 10$ now. Simulation parameters: $k_m = 5$, $\rho = 2$, $\eta = 0.3$, $v_0 = 0.4$, $m = 2$.}
\label{fig:fluctuations}
\end{figure*}

To investigate how the order parameter varies in space we choose $k_c = 0.025$ as an example and consider one scenario characterized by a relatively large noise and small density (i.e., $\eta = 0.3$ and $\rho=2$), and another one where noise is small and density large (i.e., $\eta=0.1$ and $\rho=4$). The latter is deep in the flocking regime (see snapshot Fig.~\ref{fig:model}c), while for the former the overall alignment of the cells is weaker (see snapshot Fig.~\ref{fig:model}b). The outcome of this analysis is summarized by the histograms in Figs.~\ref{fig:dist_local_op} and \ref{fig:fluctuations}, which we constructed as follows. At the end of each run we measure the order parameter $\omega_{\rm box}$ in every box of size $\mathcal{R}$ of the system. We only include the order parameter in the histogram if this box is not empty, i.e., contains at least one L or R cell, such that the order parameter is well defined. The histogram is then obtained from averaging over all independent runs. We find that this probability distribution is strongly peaked around $\omega_{\rm box} = 0$ and $\omega_{\rm box} = 1$, and that there is another maximum at $\omega_{\rm box} = 0.5$. The relatively non-monotonic structure of the distribution can be explained by some values of $\omega_{\rm box} = 0$ being much more likely to occur if they are rational numbers for ratios of small number of cells. This is particularly evident for small densities where only few particles are present in some boxes. However, in either case we find the general trend of the distribution decreasing away from $\omega_{\rm box} = 0.5$, and then strongly increasing at the edges. A noticeable difference between the two cases shown in Fig.~\ref{fig:dist_local_op} is that for higher densities and lower noises the relative frequency of the homochiral state is much greater. We have also considered an intermediate state of low noise and density ($\eta = 0.1$ and $\rho = 2$) (not shown) and did not find a significant difference from the histogram in Fig.~\ref{fig:dist_local_op}a. 

In Figs.~\ref{fig:fluctuations}a,b we present the probability distribution of the field $n_R$, defined as the number of R cells in a given box. As the distribution of L and R cells are identical, we only show one of the two. Again, we consider the average over independent runs at a fixed time where the average over all system has reached a steady state to find the probability distributions. Note that if all cells were distributed homogeneously in space, $\langle n \rangle = (2\mathcal{R})^2 \rho$. We normalize the number of cells by this number and write the renormalized quantities with a tilde, i.e., $\tilde{n}_R = n_R/\langle n \rangle$. For $k_c = 0$, i.e., if the system is homochiral, there is about a $50\%$ probability that its entire population consists of L cells, thus $P\left(\tilde{n}_R = 0\right) \approx 0.5$ (inset in Fig.~\ref{fig:fluctuations}a). Since the probability distribution of $\tilde{n}_R$ is trivial in this case we now consider only the systems that evolve to R homochirality. The distribution of $\tilde{n}_R$ is rather broad, see Fig.~\ref{fig:fluctuations}a, with a mean of $\langle \tilde{n}_R\rangle \approx 0.9$ cells per box with relative fluctuations of $\sigma\left(\tilde{n}_R\right)/\langle \tilde{n}_R\rangle \approx 0.62$, where $\sigma\left(\tilde{n}_R\right)$ is the standard deviation of the distribution. Note that the distribution peaks at $\tilde{n}_R = 0.75$ and decreases for both smaller and larger values. If we consider the total number of cells, without regard for their chirality, that is $\tilde{n} = \tilde{n}_R + \tilde{n}_L$, we find the distribution shown in Fig.~\ref{fig:fluctuations}b. It is essentially identical to the one for R cells (after removing the subset of systems that evolved to L homochirality), reflecting the fact that, after a monolayer has reached a homochiral state, cells are either all R or all L, thus the distribution of the number densities of the two sub-populations is equal to the distribution of the entire population. This is markably different for non-vanishing $k_c$. Again, we choose $k_c = 0.025$ as an example. The most likely case now is to encounter a box that contains no R cells and the distribution is monotonically decreasing for increasing $\tilde{n}_R$, see Fig.~\ref{fig:fluctuations}c. The average $\langle \tilde{n}_R \rangle \approx 0.46$ is about half the previous average value (reflecting that the mean global order parameter is $\omega = 0.5$), but the distribution is much wider, with the standard deviation almost being equal to the mean, $\sigma\left(\tilde{n}_R\right)/\langle \tilde{n}_R\rangle \approx 0.99$. Thus, fluctuations are very large. The distribution for the total number of cells in this case (Fig.~\ref{fig:fluctuations}d) is similar to the one for vanishing rate $k_c$, with $\langle \tilde{n}\rangle \approx 0.9$ and $\sigma\left(\tilde{n}\right)/\langle \tilde{n}\rangle \approx 0.75$. However, the distributions of $\tilde{n}$ and $\tilde{n}_{R}$ (or, equivalently, of $\tilde{n}_{L}$) are now different, as can be seen by comparing Fig.~\ref{fig:fluctuations}c and Fig.~\ref{fig:fluctuations}d. With increasing $k_c$, the distribution preserves its structure, but $\langle \tilde{n}_R\rangle$ slightly increases (to $\langle \tilde{n}_R\rangle \approx 3.7$ for $k_c = 10$), while the relative fluctuations slightly decrease (see inset of Fig.~\ref{fig:fluctuations}c). Mean and relative fluctuations for the total number of cells $\tilde{n}$ remains constant. For $k_c = 0.025$, but higher density and lower noise ($\rho = 4$ and $\eta = 0.1$ compared to $\rho = 2$ and $\eta = 0.3$ as before) the distribution becomes less broad, with $\langle \tilde{n}_R\rangle \approx 0.84$ and $\langle \tilde{n}\rangle \approx 0.61$, but its mean and structure do not change much.

To complete our analysis of spatial fluctuations in model cellular flocks, we look at the number density correlation functions: i.e. $\mathcal{C}_{XY} = \left\langle n_X \left(\bm{r}\right) n_Y \left(\bm{r}^\prime\right) \right\rangle$, where $X$ and $Y$ are any combination of R and L. This is shown for two different $k_c$ values -- i.e. $k_c = 0.025$ and $k_c = 10$ -- in Figs.~\ref{fig:fluctuations}e and Fig.~\ref{fig:fluctuations}f, respectively. For $k_c = 0.025$, we find that, whether R or L, like-chiral cells are more strongly correlated in space than cells of opposite chirality, even at long distances, see Fig.~\ref{fig:fluctuations}e. Furthermore, the correlation functions roughly follow a power-law decay, with the exponent associated with like-chiral cells being approximatively twice that of cells of opposite chirality. That is, for $k_c = 0.025$, $\mathcal{C}_{{\rm R}{\rm R}} \approx \mathcal{C}_{{\rm L}{\rm L}} \sim |\bm{r} - \bm{r^\prime}|^{-0.2}$, while $\mathcal{C}_{{\rm R}{\rm L}} \sim |\bm{r} - \bm{r^\prime}|^{-0.11}$, see inset in Fig.~\ref{fig:fluctuations}e. For higher density and lower noise we find that the behavior is similar, with the ratio of the exponents being again about two. For higher $k_c$, the two correlation functions instead overlap, with $\mathcal{C}_{{\rm R}{\rm R}} \approx \mathcal{C}_{{\rm L}{\rm L}} \sim |\bm{r} - \bm{r^\prime}|^{-0.18}$ and $\mathcal{C}_{{\rm R}{\rm L}} \sim |\bm{r} - \bm{r^\prime}|^{-0.16}$, see Fig.~\ref{fig:fluctuations}f. Furthermore, in this case the large length-scale behavior of the correlation functions is very similar as well.

In conclusions, our analysis revealed a large inhomogeneity in both the cell number density and chiral order parameter, despite chirality not affecting directly cellular motion. That is, Eqs.~\eqref{eq:vicsek}, which govern the motion as well as the orientational interactions among cells, do not distinguish between R and L. The higher spatial correlation of like-chiral cells is, therefore, is indicative of an emergent feedback mechanism, which effectively enhances the interactions between the like-chiral cells.

\subsection{Time to homochirality}

\begin{figure*}
\centering
\includegraphics[width=\textwidth]{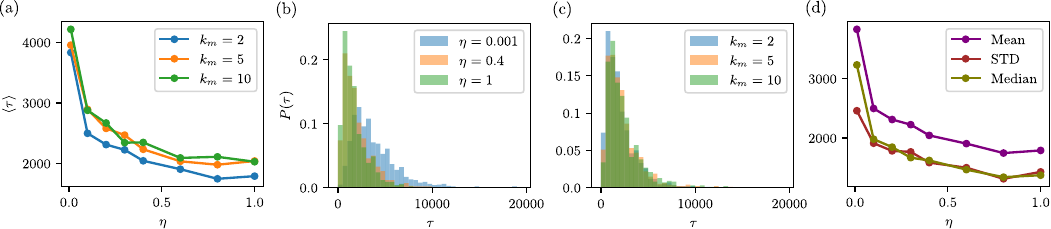}
\caption{{\bfseries Transition time for different noises and division rates.} (a) The average time $\langle \tau \rangle$ in which the system evolves for a homochiral state as a function of noise $\eta$ for different cell division rates $k_m$. (b) The distribution $P(\tau)$ of the transition time $\tau$ found from recording $\tau$ for 1000 independent runs for three values of noise. (c) The same distribution but now for three different division rates. (d) The mean (purple), standard deviation (STD, red), and median (olive green) of the distribution $P(\tau)$ for different values of noise and for one division rate, $k_m = 5$. Simulation parameters: $k_c = 0$, $\rho = 2$, $v_0 = 0.4$, $m = 2$.} 
\label{fig:Noise_RR}
\end{figure*}

In this section we investigate how cell motion affect the convergence to homochirality. To this end, we set $k_c = 0$, to guarantee that either one of the two available homochiral states are reached in a finite time $\tau$, and we reconstruct the statistics of $\tau$ for various parameter choices. Specifically, we again run $10^{3}$ independent simulations, terminating each run once homochirality is established.

We begin this analysis with an assessment of the influence of rotational noise, by varying its standard deviation $\eta$ throughout the unit interval. We find that the mean time $\langle \tau \rangle$ required to reach the homochiral state is significantly larger than for lower noises, quickly decreases, and eventually plateaus as noise is increased, see Fig. \ref{fig:Noise_RR}a. Orientational noise, thus, facilitates the onset of homochirality, by favoring the dispersion of the cells in the solvent, hence the uptake of nutrients, which is instrumental to their division. In the limiting case of vanishing noise, on the other hand, the two cells comprising the initial configuration of the system move on a straight trajectory, which becomes quickly depleted of solvent particles, thus reducing the performance of the reactions in Eqs.~\eqref{eq:Reactions}, which lead to homochirality. This effect, however, is at play only at low noise, where the dispersion of the cells in the solvent does not completely disrupt the coherence of the flock. For large $\eta$ values, conversely, the monolayer transitions from flocking to isotropic and the route to homochirality is equivalent to that of isolated cells. To clarify this further we show in Fig.~\ref{fig:Noise_RR}b the probability distributions of $\tau$ for a few different $\eta$ value. We find that noise has no visible influence on the mode of the distribution (i.e. the location of the peak), but does increase the length of its tail. In conclusion, flocking slows down the convergence to homochirality, by reducing the mixing of the cells, hence the effective biological noise. However, such an effect can be drastically reduced by introducing a small amount of orientational noise, which can restore an efficient mixing without disrupting the flock.

Next, we consider different values for the cell division rate $k_m$. Remember that the rate $k_m$ is defined relative to the death rate such that $k_m > 1$ is required for a growing cell population (greater division than death rate) and that for $k_m < 1$ the cells in a given system will eventually all die. Surprisingly, the value of the ratio is rather irrelevant, with the mean length for $k_m = 2$ and $k_m = 10$ being very similar even though in the former case the cells are dying at a rate five times higher, see Fig. \ref{fig:Noise_RR}a. In particular, different rates show the same power law behavior. To illustrate the similarity of the three different ratios we consider, we present in Fig. \ref{fig:Noise_RR}c the probability distribution for different rate ratios at a fixed noise. Indeed, they are almost indistinguishable. We find that these probability distributions are again strongly peaked at small times, but that there is a long tail with some runs taking almost five times the average time to reach homochirality. To quantify the probability distribution of the time to homochirality we present the mean, standard deviation, and median for a fixed rate ratio $k_m = 5$ for different noises. Note that the results for $k_m = 2$ and $k_m = 10$ is almost identical. We find that all three quantities have a similar magnitude and fall off at a similar rate, with the standard deviation and mean curves overlapping while the mean is shifted by a constant factor with respect to these curves, see Fig.~\ref{fig:Noise_RR}d. All three curves follow the same power-law behavior $\sim \eta^{-0.16(\pm0.03)}$.

\begin{figure*}
\centering
\includegraphics[width=\textwidth]{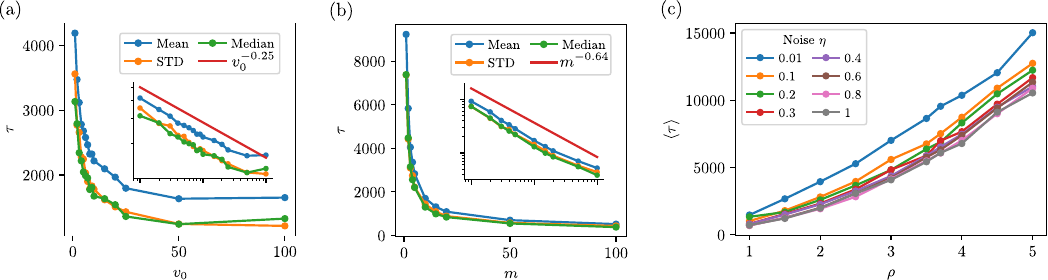}
\caption{{\bfseries Transition time for different speed and density} (a) The mean (blue), standard deviation (STD, orange), and median (green) of the distribution $P(\tau)$ for different values of the speed $v_0$ of the Vicsek model. The inset shows the same data but in a log-log plot, with the red line indicating the power law decay of all three quantities  $\sim v_0^{-0.25(\pm 0.02)}$. (b) The data as in panel (a) but when varying $m$, the number of steps of the reactions performed during each step of the Vicsek model. Mean, standard deviation, and median now decay as $\tau \sim m^{-0.64(\pm0.02)}$. (c) The average time $\langle \tau \rangle$ as a function of the density $\rho$. Each curve corresponds to a different values of noise $\eta$. Simulation parameters: $k_c = 0$, $k_m = 5$, $\rho = 2$ (in panels (a) and (b)), $v_0 = 0.4$ (in panels (b) and (c)), $m = 2$ (in panels (a) and (c)).}
\label{fig:speeds}
\end{figure*}

We continue by looking at the effect of the cell speed $v_0$, as well as the number $m$ of cycles of the Gillespie algorithm used to update the stochastic trajectory arising from Eqs.~\eqref{eq:Reactions} in one time step of the Vicsek model. For both parameters we perform an analysis similar as the one for the noise presented above. Varying the speed $v_0$ we find that $\tau$ decreases with increasing speed, see Fig.~\ref{fig:speeds}a. The decrease is fast at low speed and slower for large $v_0$ values. The higher the speed the more peaked the distribution at small times. Again, we find that standard deviation and median curves are very similar, while the mean $\langle \tau \rangle$ changes only by a pre-factor, so that all three quantities exhibit the same power-law scaling (inset in Fig.~\ref{fig:speeds}a). A similar behavior is found when varying $m$. The higher the number of cycles the faster the convergence to homochirality, see Fig. \ref{fig:speeds}b. The most significant difference compared with the speed $v_0$ is that the decay is much steeper, see Fig. \ref{fig:speeds}b.

Finally, we explore the effect of density $\rho$ by varying the number of cells and solvent particles, while keeping the magnitude of noise fixed. Together with $\eta$, density is a classical control parameter of the Vicsek model, which determines whether the system is in the isotropic or flocking phase. Upon increasing $\rho$, we find that the mean time increases slightly faster than linear for all $\eta$ values. Away from the smallest noise value $\eta = 0.01$, we do not find a significantly different behavior when varying the noise at a fixed density, see Fig.~\ref{fig:speeds}c. Lastly, we note that when increasing $\mathcal{R}$, the mean time increases approximately linearly with $\mathcal{R}$, however the value of the noise becomes less important, with the time being considerably less sensitive to changes in noise, as expected since the alignment interaction radius is increased. 

\section{\label{sec:discussion}Discussion}

For sake of completeness, we briefly review here the theory of Jafarpour {\em et al.} and discuss how the mechanism presented in Refs.~\cite{jafarpour2015, jafarpour2017}, used explain the onset of molecular homochirality, combined with our findings, allows one to formulate a possible explanation of the chiral cellular flows investigated by Duclos {\em et al}.~\cite{duclos2018} and others~\cite{hoffmann2020,yashunsky2022}, and also to sketch a possible generic route to the establishment of spatially uniform traits in cellular flocks.

To account for spatially extended systems, Jafarpour {\em et al.} coupled Eqs.~\eqref{eq:Reactions} with a diffusive dynamics, obtaining the following reaction-diffusion equation for the space-dependent order parameter:
\begin{equation}\label{eq:StochasticDiffEq}
\frac{d\omega}{dt} = - \frac{2 V k_c k_d}{N k_m} \left(\omega-\frac{1}{2}\right) + \mathcal{D} \nabla^2\omega + \sqrt{\frac{2 k_d}{N} \omega \left(1 - \omega\right)}\,\zeta\;.
\end{equation}
Here, $N \gg 1$ is the number of agents -- whether molecules, cells or other -- $V$ the volume of the system, $\mathcal{D}$ a diffusion coefficient, and $\zeta$ a Gaussian white noise of zero mean and unit variance. Both in Refs.~\cite{jafarpour2015,jafarpour2017} and here, the noise field $\zeta$ is independent of the stochastic processes affecting the motion of the agents, but reflects the inherent noise of the reactions governing the inheritance of chirality or other traits. Eq.~\eqref{eq:StochasticDiffEq}, in turns, allows a simple explanation of the origin of homochirality in the limit of vanishing $k_c$. If the system is not spatially extended, diffusion is irrelevant and Eq.~\eqref{eq:StochasticDiffEq} reduces to
\begin{equation}\label{eq:fixed_points}
\frac{d \omega}{dt} =  \sqrt{\frac{2 k_d}{N} \omega \left(1 - \omega\right)}\,\zeta\;,
\end{equation}
when $k_c = 0$. This equation has two fix points -- i.e. $\omega = 0$ and $\omega =1$ -- representing the two homochiral states. Perhaps more interestingly, the right-hand side of Eq.~\eqref{eq:fixed_points} is coupled to the noise field $\zeta$, indicating that noise is indispensable for the onset of homochirality. As long as $k_c=0$, diffusion does not change this picture, as the structure of the fixed points is not altered by the Laplacian term in Eq.~\eqref{eq:StochasticDiffEq}.

The cellular processes described by Eqs.~\eqref{eq:Reactions} are formally identical to those considered by Jafarpour {\em et al.} The spatial dynamics of the agents is, however, completely different and cannot be treated with the same approach of Refs.~\cite{jafarpour2015,jafarpour2017}, which leads to Eq.~\eqref{eq:StochasticDiffEq}. Although deriving a field equation analogous to Eq.~\eqref{eq:StochasticDiffEq} is outside the scope of this article, it is still possible to rationalize the occurrence of homochirality in the regime where $k_c=0$ as follows. In the limit of $\eta \to 1$, the system is in the isotropic phase and the previous picture applies without variations of any sort. By contrast, in the flocking regime, but close to the isotropic-flocking transition, the anisotropy associated with the spatial dynamics of the agents can be effectively accounted for by means of an anisotropic diffusion tensor, so that $\mathcal{D}\nabla^{2}\omega\to\partial_{i}(\mathcal{D}_{ij}\partial_{j}\omega)$. This, again does not change the structure of the fixed points, thereby guaranteeing the existence of homochirality. Well in the flocking state, this reasoning is no longer valid. Yet our numerical evidence suggests that the same phenomenology persists also in this regime, although the reduced mixing of the agents slows down the convergence to homochirality.
\begin{figure}
\centering
\includegraphics[width=0.75\columnwidth]{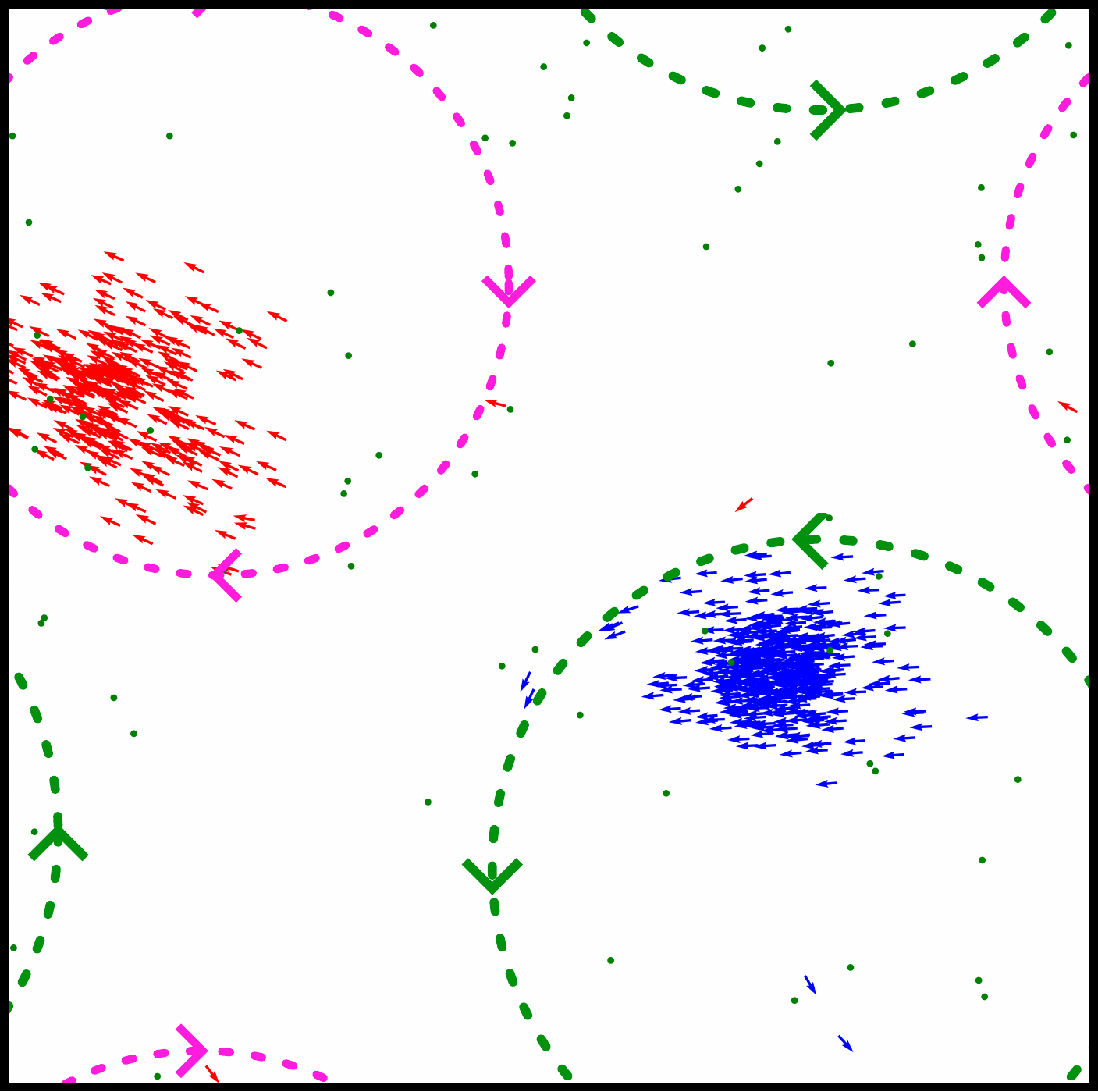}
\caption{\textbf{Separated clusters in presence of chiral motion.} Snapshot of a system with non-vanishing deterministic rotation, $\Omega = 0.1$ and $\eta = 0.01$. The system quickly reaches a state with two separated clusters that are moving on a circular trajectory. Most particles are confined to the area indicated by the dashed lines, and this state is stable for long times.}
\label{fig:Rotation}
\end{figure}

Finally, we briefly comment on the effect of including a deterministic rotation in the equation of motion, that is $\Omega \neq 0$ in Eq.~(\ref{eq:vicsek}b). In this case the chirality of a given cell explicitly modifies its spatial dynamics, and, in particular, the equations of motion now differ for L and R cells, due to their different directions of orientations, quantified by the parameter $\lambda$. Introducing this term favors flocking of cells of same-chirality since particles of different chirality now have divergent trajectories. For sufficiently large values of $\Omega$ (and sufficiently small values of $\eta$) this can result in a phase separation with flocks of different chirality coexisting in different areas of the system. The rotational motion effectively traps cells in a circular domain, resulting in a phase separated state that is stable over long times. An example is shown in the snapshot in Fig.~\ref{fig:Rotation}.

\section{Conclusion}

In this article we investigated the onset of homochirality in a population of collectively moving cells, by coupling the Vicsek 
model reaction equations modelling cell division, death, and an influx of cells from the external environment. We found that the system is guaranteed to evolve to a homochiral state from an initially symmetrically mixed state in finite time only if the system is closed in the sense that the reaction rate $k_c$ vanishes. In the mixed state we find large fluctuations of the local order parameter and the particle density around the mean value. In particular, we find that particles of same chirality tend to be correlated more strongly in space than particles of opposite chirality. In the case where the system evolves to a homochiral state, we showed that the transition time has a fat-tail distribution with ratio of mean and standard deviation being of order unity. Introducing a small amount of noise in the spatial dynamics significantly decreases the mean transition time. Lastly, we found that the time decays like a power law with the speed $v_0$ of the Vicsek model and the number of steps $m$ of the reactions.

Furthermore, while the investigation of chirality and the question how homochirality emerges was the motivation and starting point of this investigation, we note that the model we investigated is more general. Since for $\Omega=0$ the chirality of the particles does not enter the equations of motion -- i.e. the equations of motion are the same for left- and right-chiral particles -- the model can be more generally considered a model for the competition and spread of an arbitrary trait in a population of particles interacting through a Vicsek-like alignment interactions. One could therefore use this model to investigate the evolution of different properties in cellular flocks, but other biological systems can also be considered. For example, the spatial dynamics of some types of bacteria have been successfully described using the Vicsek model (see, e.g., Refs.~\cite{czirok1996,gregoire2001,ginelli2016,nishiguchi2017,holubec2021}). Our model (or a potentially slightly modified version) can therefore be used to describe the evolution of chirality, inhomogeneous phenotypes, motility, or other properties in bacterial colonies. Furthermore, bacterial colonies often consist of several interacting species, and our model can be applied to the study of the dynamics of such systems. These examples connect our model with the recent work of, for example, Refs.~\cite{zuo2020,peled2021,jose2022,chatterjee2023}. In particular, experimental realizations using bacteria colonies might be more accessible than cellular systems. On the other hand, there is evidence that the chirality of cells influences their spatial dynamics in that the left- and right-handed cells move differently in space \cite{taniguchi2011,duclos2018,hoffmann2020,yashunsky2022}. Thus, a natural extension of the model above is to extend it to account for chirality-dependent spatial dynamics. We have briefly commented on the effect of $\Omega \neq 0$ on the flocking, but this was not the focus of the present work, and a closer analysis is necessary. Finally, an analytical treatment of the model numerically investigated here could be developed following Jafarpour {\em et al.} \cite{jafarpour2015, jafarpour2017}. A stochastic differential equation for the order parameter can be derived from the reaction rates. The main difficulty of generalizing this approach to our model consists in including the anisotropic, non-equilibrium spatial dynamics. A starting point for an analytical model could, for example, be the theory of ``Malthusian flocks'', where the number of flocking entities is not conserved~\cite{toner2012}.

While framed in terms of chirality of cellular flocks, the model introduced above is therefore more general. The novel coupling of reaction rates with the Vicsek model described in this work allows for numerous extensions. It can be used to investigate the evolution and spreading of biological traits (not necessarily of binary nature like chirality) in flocking systems, as well as systems consistent of multiple interacting species.

\section*{Acknowledgements}

This work was supported by the Netherlands Organization for Scientific Research (NWO/OCW), as part of the Vidi scheme (L.A.H. and L.G.), and by the European Union via the ERC-CoGgrant HexaTissue (L.G.) L.A.H. thanks Livio Nicola Carenza, Ireth Garc\'ia-Aguilar, and Victor Yashunsky for valuable discussions. Part of this work was performed using the ALICE compute resources provided by Leiden University.

\bibliography{Biblio.bib}

\end{document}